\documentclass[letterpaper,10pt]{article} 

\usepackage{osameet3} 

\usepackage{amsmath,amssymb, amsfonts, amsthm, bm}
\usepackage{newtxmath}
\usepackage{commath}
\usepackage[colorlinks=true,bookmarks=false,citecolor=blue,urlcolor=blue]{hyperref} 

\usepackage{lipsum}
\usepackage[mode=tex]{standalone} 

\usepackage{subfigure}
\usepackage{dsfont}
\usepackage{float}
\usepackage{algorithm}
\usepackage[noend]{algpseudocode}
\usepackage{tikz,pgfplots}
\usepgfplotslibrary{fillbetween}
\usepgfplotslibrary{groupplots}
\usetikzlibrary{shapes.geometric}

\usepackage{paralist}
\usepackage{xcolor}
\tikzset{%
	partial ellipse/.style args={#1:#2:#3}{%
		insert path={+ (#1:#3) arc (#1:#2:#3)}%
	}%
}%

\makeatletter
\def\BState{\State\hskip-\ALG@thistlm}
\makeatother

\begin{document}
\title{ End-to-end Autoencoder for Superchannel Transceivers with Hardware Impairment}

\vspace{-0.6cm}
\author{
    Jinxiang Song\textsuperscript{(1)},  Christian H\"{a}ger\textsuperscript{(1)}, 
     Jochen Schr\"{o}der\textsuperscript{(2)},\\  Alexandre Graell i Amat\textsuperscript{(1)},  
    and Henk Wymeersch\textsuperscript{(1)}
}
\address{\textsuperscript{(1)} Department of Electrical Engineering, Chalmers University of Technology, Gothenburg, Sweden\\
   \textsuperscript{(2)} Department of Microtechnology and Nanoscience, Chalmers University of Technology, Gothenburg, Sweden\\
    }
\email{jinxiang@chalmers.se}

\vspace{-0.7cm}
\begin{abstract}
We propose an end-to-end learning-based approach for superchannel systems impaired by non-ideal hardware component. Our system achieves up to $60\%$ SER reduction and up to $50\%$ guard band reduction compared with the considered baseline scheme.

\end{abstract}

\vspace{0.0cm}
\section{Introduction}

Modern high spectral efficiency (SE) superchannel systems require dense channel spacing (close to the symbol rate) coupled with high-order modulation formats.
The narrow channel spacing and high-order modulation formats,  however, exacerbate performance degradation due to non-ideal hardware components~\cite{5621166, liu2013joint}.
To compensate for the hardware impairments, various digital pre-distortion (DPD) algorithms
including arcsin~\cite{curri2012optimization, tang2008coherent} and Volterra series-based pre-compensation~\cite{khanna2015robust, berenguer2015nonlinear,
elschner2018improving,yoffe2019low} have been proposed.

As an alternative to these conventional approaches, deep learning (DL) techniques using neural networks (NNs) have recently been proposed for hardware impairment compensation in both wireless ~\cite{benvenuto1993neural,tarver2019design, gotthans2014digital,tarver2019neural}  and optical communications~\cite{paryanti2018recurrent,schaedler2019ai, abu2019neural, paryanti2020direct, Bajaj2020}, 
where the main idea is to replace the DPD block with an NN that can be optimized from data.
In contrast to focusing on a specific functional block (e.g., DPD), end-to-end learning using an autoencoder (AE) optimizes the transmitter and receiver jointly\cite{o2017introduction}. However, the methods in \cite{benvenuto1993neural, gotthans2014digital, tarver2019design, tarver2019neural,paryanti2018recurrent,schaedler2019ai, abu2019neural, paryanti2020direct, Bajaj2020} are derived considering a single-channel setup, thus ignoring inter-channel interference (ICI) 
with neighboring channels, which can yield significant performance degradation in densely-spaced superchannel systems.


\textcolor{black}{In this paper, we propose a novel end-to-end AE for superchannel systems impaired by non-ideal hardware component, the nonlinear IQ-Modulator (IQM) in particular, that limits ICI.} Simulation results show that our method
achieves significantly better symbol error rate (SER) performance than a conventional baseline scheme. Moreover, the proposed approach shows great potential to increase the SE of superchannel systems by reducing the guard band with limited impact on the SER.


\vspace{-0.1cm}
\section{System model}

\begin{figure}[b]
    \vspace{-0.3cm}
    \centering
    \tikzstyle{expr}=[font=\footnotesize]%
\tikzstyle{block}=[expr,rectangle, draw, thick, minimum size=30pt,
minimum height=17pt, inner sep=3pt, rounded corners=5, fill=white]%
\tikzstyle{edge} = [draw, thick, ->]%
\begin{tikzpicture}[scale=1.2, auto, swap, every text node part/.style={align=center}]%

	\def\xoff{0.45};%
	\def\yoffa{0.35};%
	\def\yoffb{1.1};%
	
	\draw[black, dashed, rounded corners=5, fill=black!5] (-8.9, \yoffa+1.6) rectangle (0.5, \yoffb-1.2);%
	\draw[black, dashed, rounded corners=5, fill=black!5] (-9.0, \yoffa+1.2) rectangle (0.4, \yoffb-1.7);%
	\draw[black, dashed, rounded corners=5, fill=black!5] (-9.1, \yoffa+0.8) rectangle (0.3, \yoffb-2.2);%
	
	\node[] (txn)           at (-6.8, \yoffa+1.4) { Tx-N};
	\node[rotate=50] (dots) at (-7.3, \yoffa+1.2) { \large{$\cdots$}};
	\node[] (tx2)           at (-7.5, \yoffa+1.0) { Tx-2};
	\node[] (tx1)           at (-8,   \yoffa+0.6)    { Tx-1};

    \node[] (msg1)               at (-9.9, \xoff) {};%
	\node[block,] (tx1)          at (-8.5, \xoff) {Mapper};%
	\node[block, rotate=0] (os1) at (-7.1, \xoff) {$\uparrow$};%
	\node[block] (tps1)          at (-5.5, \xoff) {pulse shaping};%
	\node[block, fill=black!15](arcsin1) at (-3.9,\xoff) {DPD};
	
	\path[edge] (msg1) -- node[above, pos=0.3]{$m_k$} (tx1);%
	\path[edge] (tx1)  -- node[above, pos=0.4]{$x_k$}  (os1);%
	\path[edge] (os1)  -- node[above, pos=0.5]{$\tilde{x}_k$} (tps1);%
	\path[edge] (tps1) -- node[above, pos=0.4]{$s_k$}(arcsin1);%
	
	\path[draw, thick, -] (-3.45,\xoff) -- (-3.2,0);

	\node[] (msg)                   at (-9.9, -\xoff) {};%
	\node[block, fill=yellow] (tx)  at (-8.5, -\xoff) {NN1};%
	\node[block,rotate=0] (os)      at (-7.1, -\xoff) {$\uparrow$};%
	\node[block, fill=yellow] (tps) at (-5.5, -\xoff) {NN2};%
	
	\node[block, fill=yellow](arcsin) at (-3.9,-\xoff) {NN3};
	
	\path[edge] (msg) -- node[above, pos=0.3]{$m_k$} (tx);%
	\path[edge] (tx)  -- node[above, pos=0.4]{$x_k$} (os);%
	\path[edge] (os)  -- node[above, pos=0.3]{$\tilde{x}_k$}(tps);%
	\path[edge] (tps) -- node[above, pos=0.6]{$s_k$}(arcsin);%
	
	\path[draw, thick, -] (-3.45,-\xoff) -- (-3.2,0);
	
	\node[block] (dac) at (-2.4, 0) {DAC};%
	\node[]  at (-1.3, 0.4) {linear};%
	\node[block] (pa) at (-1.3, 0) {PA};%
	
	\path[edge] (-3.2,0) -- node[above, pos=0.4]{$\tilde{s}_k$} (dac);
	\path[edge] (dac) -- (pa);
	\node[block] (mod) at (-0.2, 0) {IQM};%

	\node [trapezium, trapezium angle=45, minimum width=20mm, draw, thick, rotate=270] (mux) at (1.2,0.3) {\,\,\,\,MUX\,\,\,\,};
	\path[edge] (0.5, 0.6) --  (1,0.6);
	\path[edge] (0.4, 0.3) --  (1,0.3);
	\path[edge] (mod) --  (1,0);

	\node[expr, circle, draw, inner sep=-0.3] (plus) at (2.2, 0.3) {$+$};%
	\node[] (noise) at (2.2, -0.7) {$n(t)$};
	
    \node [trapezium, trapezium angle=45, minimum width=20mm, draw, thick, rotate=90] (demux) at (3.2,0.3) {DEMUX};
	\path[edge] (3.4, 0.6) --  (4.4, 0.6);
	\path[edge] (3.4, 0.3) --  (4.3,0.3);

	\node[] (receiver) at (6,1.6) {Receiver};
	\draw[black, dashed, rounded corners=5, fill=black!5] (4.4, \yoffa+1.6) rectangle (10.95, \yoffb-1.8);%
	\draw[black, dashed, rounded corners=5, fill=black!5] (4.3, \yoffa+1.2) rectangle (10.85, \yoffb-2);%
	\draw[black, dashed, rounded corners=5, fill=black!5] (4.2, \yoffa+0.8) rectangle (10.75, \yoffb-2.2);
	
	\node[] (rx1) at (6.7,\yoffa+1.4) { Rx-N};
	\node[rotate=50] (dots) at (6.2,\yoffa+1.2) { \large{$\cdots$}};
	\node[] (rx2) at (6.0,\yoffa+1.0) { Rx-2};
	\node[] (rx3) at (5.5,\yoffa+0.6) { Rx-1};

    \node[block, rotate=0] (lpf) at (5.0, 0) {ADC};%

	\node[block] (mf) at (6.2, 0) {MF};%
	\node[block, rotate=0] (ds) at (7.5, 0) {$\downarrow$};%

	\path[draw, thick, -] (ds) --node[above, pos=0.6] {$y_k$}  (8.3,0);%

	\node[block] (rx1) at (9.3, \xoff) {De-mapper};%
	\node[block, fill=yellow] (rx2) at (9.2, -\xoff) {NN4};%
	
	\node[block, rotate=90] (argmax) at (10.4, -\xoff) {$\arg\max$};%

	\node[] (rmsg1) at (11.7, \xoff) {};%
	\node[] (rmsg2) at (11.7, -\xoff) {};%

	\path[edge] (pa) -- (mod);%
	\path[edge] (mux) -- (plus);%
	\path[edge] (plus) -- (demux);%
	\path[edge] (3.4,0) --  (lpf);%
	\path[edge] (noise) -- (plus);%
	\path[edge] (lpf) --  (mf);
    
    \path[edge] (mf) --(ds);%
    
    \path[edge] (rx2) --node[above, pos=0.6] {$\boldsymbol{q}_k$} (argmax);
    
    \path[edge] (8.3,0)  --  (8.65,\xoff);%
    \path[edge] (8.3,0)  --  (8.75,-\xoff);%
    \path[edge] (rx1)    -- node[above, pos=0.82]{$\hat{m}_k$} (rmsg1);%
    \path[edge] (argmax) -- node[above, pos=0.7]{$\hat{m}_k$} (rmsg2);%

    \draw[thick] (10,-\xoff) -- (10,-1.5) -- (-9.2,-1.5);
    \node[] (loss) at (2, -\xoff-0.8) {$\mathcal{J}_{CE}(\boldsymbol{\theta}_1, \boldsymbol{\theta}_2, \boldsymbol{\theta}_3, \boldsymbol{\theta}_4)$};%
    \path[edge] (-9.2,-1.5) -- (-9.2,-\xoff) ;

\end{tikzpicture}%
    \vspace{-0.2cm}
    \caption{{Block diagram showing the end-to-end system model ($\uparrow$: upsampling, $\downarrow$: downsampling). The top branch corresponds to the conventional baseline scheme and the bottom branch to the proposed AE learning-based scheme.}}
    \label{fig:system_model_wdm}
    \vspace{-0.9cm}
\end{figure}
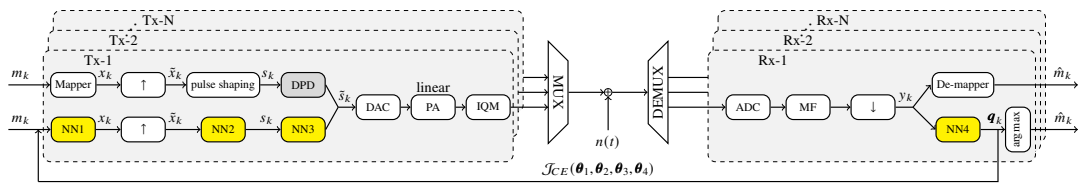
Fig.~\ref{fig:system_model_wdm} depicts the considered superchannel system. For each channel, messages belonging to the set $\mathcal{M}=\{1,\ldots,M\}$ are mapped to constellation points belonging to the set $\mathcal{X}=\{x_1,\ldots,x_M\} $. 
The upsampled signals are convolved with a pulse-shaping (PS) filter, after which a DPD algorithm is applied to pre-compensate for the IQM nonlinearity. 
The pre-compensated signal is then fed to the digital-to-analog converter (DAC) and amplified by a linear electrical power amplifier (PA) to drive the  IQM.
Similar to ~\cite{curri2012optimization,khanna2015robust, berenguer2015nonlinear}, 
we consider a back-to-back setup, where additive white Gaussian noise (AWGN) with constant power is added to simulate the  noise introduced by the booster amplifier. At the receiver, the received signals are passed through an analog-to-digital converter (ADC) and then convolved with a matched filter (MF). Finally, the downsampled signals are mapped to estimates of the transmitted messages.

\vspace{-0.1cm}
\section{AE-based superchannel system}
In principle, the entire transmitter and receiver can be implemented as an
AE and optimized by end-to-end learning as proposed in \cite{o2017introduction}. However, 
this will lead to i) a black-box solution that can be difficult to interpret  and ii) potentially high training complexity (a very complex transmitter NN with large memory may be required). To address both issues, we follow a different approach where the transmitter NN is decomposed into a concatenation of simpler NNs, each corresponding to one functional block of a conventional communication system.
By doing this, an additional advantage is that the parameters of these NNs can be initialized such that they initially perform close to their conventional counterparts. As a result, the proposed scheme has \emph{increased interpretability} and  \emph{decreased training complexity} as compared to a conventional AE. 
We note that a similar approach has recently been applied for nonlinear optical channels in ~\cite{uhlemann2020deep}, where the transmitter implementation includes a trainable symbol mapper and a trainable PS filter, and it has been show that a PS filter can be trained to compensate the chromatic dispersion and Kerr nonlinearity.



\vspace{-0.1cm}
\subsection{Proposed trainable AE}

As  shown in the bottom branch of Fig.~\ref{fig:system_model_wdm}, the AE transmitter is 
implemented as a concatenation of three NNs.  These NNs are denoted by 
$f_{\boldsymbol{\theta}_1}(\cdot)$, $f_{\boldsymbol{\theta}_2}(\cdot)$, $f_{\boldsymbol{\theta}_3}(\cdot)$
 , where $\boldsymbol{\theta}_1,  \boldsymbol{\theta}_2$, and $ \boldsymbol{\theta}_3$ 
 are the sets of trainable parameters, and are defined as follows:
\begin{compactitem}
    \item NN1 $f_{\boldsymbol{\theta}_1}$: $\mathcal{M} \to \mathbb{C}$ maps a given message $m_k\in \mathcal{M}$ to  a constellation point according to $x_k = f_{\boldsymbol{\theta}_1}(m_k)$, where an average power constraint $\mathbb{E}\{|x_k|^2\}=1$ is enforced by a normalization layer ~\cite{o2017introduction}. 
    \item NN2 $f_{\boldsymbol{\theta}_2}$: $\mathbb{C}^N \to \mathbb{C}$ generates the pulse-shaped signal 
    according to $s_{k} = \boldsymbol{\theta}_2^\top \boldsymbol{\tilde{x}}_{k}$,
    where $\boldsymbol{\tilde{x}}_{k} = [\tilde{x}_{k-N},\ldots, \tilde{x}_{k}]^\top$  is a sequence of $N$ complex-valued input signals. Note that NN2 only has a single layer applying a linear activation function, and can be interpreted as a standard finite impulse response filter.
    

    \item NN3 $f_{\boldsymbol{\theta}_3}$: $\mathbb{C} \to \mathbb{C}$ generates the pre-distorted signal according to $\tilde{s}_{k}=f_{\boldsymbol{\theta}_3}(s'_{k})$, 
    where $f_{\boldsymbol{\theta}_3}(\cdot)$ operates separately on the in-phase and quadrature branches,
    and $-1\leq \Re{\{s'_k\}, \Im{\{s'_k\}}}\leq1$
    is obtained by normalizing $s_k$ according to $s'_k = s_k/\max\{\max\{\lvert\Re\{\boldsymbol{s}\}\rvert\}, \max\lvert\Im\{\boldsymbol{s}\}\rvert\} \}$, where $\boldsymbol{s}$ is the pulse-shaped signal sequence and $\lvert \Re\{\boldsymbol{s} \}\rvert$ and $\lvert \Im\{\boldsymbol{s} \}\rvert$  return the absolute value of the real and imaginary part of each element in $\boldsymbol{s}$, respectively.
    
\end{compactitem}
At the receiver, NN4 $f_{\boldsymbol{\theta}_4}$: $\mathbb{C} \to \mathcal{M}$ maps the downsampled signal $y_k$  to an $M$--dimensional  probability vector according to $\boldsymbol{q}_k=f_{\boldsymbol{\theta}_4}(y_k)$,  where $\boldsymbol{\theta}_4$ is the set of trainable parameters. The transmitted message is estimated according to $\hat{m}_k=\arg\max_m[\boldsymbol{q}_k]_{m}$, where $[\boldsymbol{x}]_m$ returns the $m$-th element of $\boldsymbol{x}$. 

\vspace{-0.1cm}
\subsection{Optimization procedure }
The system is trained in an end-to-end manner by minimizing the cross-entropy loss defined by $\mathcal{J}_\text{CE}(\boldsymbol{\theta}_1,\boldsymbol{\theta}_2, \boldsymbol{\theta}_3,\boldsymbol{\theta}_4)=\mathbb{E}\{ \log [f_{\boldsymbol{\theta}_4}(y_k)]_{m_k}\}$, where the dependence of $\mathcal{J}_\text{CE}(\boldsymbol{\theta}_1,\boldsymbol{\theta}_2, \boldsymbol{\theta}_3,\boldsymbol{\theta}_4)$  on $\boldsymbol{\theta}_1,\boldsymbol{\theta}_2, \boldsymbol{\theta}_3$ is implicit through the distribution of the downsampled signal $y_k$, which is a function of the channel input
$g(\tilde{s}_k)$,
where $g(\cdot)$ denotes the transfer function of the DAC, PA and IQM, and $\tilde{s}_k$ is dependent on NNs 1--3 as can be seen in the bottom branch of Fig.~\ref{fig:system_model_wdm}. In practice, $\mathcal{J}_\text{CE}$ can be approximated via Monte Carlo simulation according to $\mathcal{\hat{J}}_\text{CE}(\boldsymbol{\theta}_1,\boldsymbol{\theta}_2, \boldsymbol{\theta}_3,\boldsymbol{\theta}_4) \approx  \frac{1}{B_\text{s}}\sum_{k=1}^{B_\text{s}} \log [f_{\boldsymbol{\theta}_4}(y_k)]_{m_k}$, where $B_\text{s}$ is the mini-batch size.
For the optimization, in order to have a faster and more stable convergence, the NNs are first initialized to mimic their model-based counterparts via pre-training. Then, the sets of parameters $\boldsymbol{\theta}_1,\boldsymbol{\theta}_2, \boldsymbol{\theta}_3,\boldsymbol{\theta}_4$ are jointly optimized using the Adam optimizer.
\label{opt_procedure}


\vspace{-0.1cm}
\section{Results}
We set $M=64$ and consider a 3-channel system where the guard band between the adjacent channels is $\eta f_{\mathrm{b}}$, where  $\eta\geq 0$ and $f_{\mathrm{b}}$ is the symbol rate. 
\textcolor{black}{The hardware impairment considered in this paper is restricted to the IQM nonlinearity, while it should be noted that the proposed approach can be straightforwardly applied to a more general setup where the other transmitter components are not idealized.} As in~\cite{curri2012optimization, tang2008coherent}, we consider the IQM transfer function $E_\text{out}=\sin{(E_{\text{in}}\, \pi/2)}$, where $E_\text{in}$ is the driving signal with a peak voltage $V_\text{p}$.
At the receiver, the ADC is implemented as a brick-wall filter (i.e., with $2f_{\mathrm{b}}$ bandwidth) followed by a sampler with rate $2f_{\mathrm{b}}$, and the MF is fixed to a root-raised cosine (RRC) filter, which is also used for PS in the baseline. Furthermore, for the baseline,  we use a geometrically-shaped constellation set obtained by training a standard AE~\cite{o2017introduction} over the AWGN channel at $\text{SNR} = 18\, \mathrm{dB}$, and apply the arcsin with clipping based DPD~\cite{curri2012optimization},  according to $\tilde{s}_k = \min\{V_{\text{clip}}, \arcsin (s_k)\}$, where $V_\text{clip}$ is referred to as the clipping factor and $\arcsin(\cdot)$ operates separately on in-phase and quadrature branches, to compensate the IQM nonlinearity. The system performance is evaluated by measuring the achieved SER.  
\label{baseline}

\begin{figure}[t]
    \centering
    \vspace{-0.45cm}
    \begin{tikzpicture}

\begin{semilogyaxis}[%
    scale only axis,
	grid=both,%
	xlabel=$V_{\text{p}}$,
	ylabel= symbol error rate,
	xmin=.6, xmax=1.0,%
	ymin=1e-6, ymax=1e-1,%
 	width=6.5cm, height=3.8cm,%
	legend cell align={left},%
	label style={font=\footnotesize},
	tick style ={font=\footnotesize},
	legend style={at={(0.55,0.85)},anchor=west,nodes={scale=0.9, transform shape},font=\footnotesize}%
]%

\addplot +[thick, color=blue, solid, mark=*, mark options={scale=1,solid }] coordinates { 
(0.3, 0.12709875) (0.4, 0.03511906) (0.5, 0.00793437)
(0.6, 0.00188344 ) (0.7,  0.00056078) (0.8, 0.00045375)
 (0.9, 0.00116547) (1.0, 0.00255531)
};
\addlegendentry{W/o compensation}

\addplot +[thick, color=green!50!black, solid, mark=square, mark options={scale=1,solid }] coordinates { 
(0.3, 2.36095156e-01) (0.4, 9.16050000e-02 ) (0.5, 2.98837500e-02 )
(0.6,  8.01875000e-03 ) (0.7,  1.90562500e-03 ) (0.8,4.16093750e-04 )
 (0.9, 8.23437500e-05) (1.0,3.25000000e-05])
 (1.1, 1.50781250e-04)
};
\addlegendentry{Baseline}

\addplot +[thick, color=red, solid, mark=triangle*, mark options={scale=1,solid }] coordinates { 

(0.6, 5.95e-5 ) (0.7,  1.52e-5 ) (0.8,3.2e-6 )
 (0.9, 1.69e-6) (1.0,3.78e-6])
 
};
\addlegendentry{AE}
\end{semilogyaxis}

\node at (0.7, 3.6) {(a)};
\end{tikzpicture}

\begin{tikzpicture}

\begin{axis}[
    scale only axis,
    xlabel=Normalized frequency $(f/f_{\text{b}})$,
    ylabel= Frequency response $|H(f)|$,
	xmin=-2, xmax=2,%
	ymin=-0.1, ymax=1.7,%
 	width=6.5cm, height=3.8cm,%
	legend cell align={left},%
	label style={font=\footnotesize},
	ylabel style={font=\footnotesize, at={(0.05,0.5)}},
	tick style={font=\footnotesize},
	legend style={at={(0.26,0.3)},anchor=west,nodes={scale=0.8, transform shape}, font=\footnotesize}%
]

\addplot+[thick, color=green!50!black, solid, mark=None,line width=0.5mm] table {Figures/Data/rrc_middle.txt};
\addlegendentry{RRC filter}
\addplot+[thick, color=black, dashed, mark=None, mark options={scale=1,solid }] table {Figures/Data/rrc_1.05.txt};
\addlegendentry{RRC adj. ch., $\eta = 0.05$}

\addplot+[thick, color=red, mark=None,line width=0.3mm] table {Figures/Data/learned_fr.txt};
\addlegendentry{Learned filter, $\eta=0.05 $}


\addplot+[thick, color=blue, dashed, mark=None, line width=0.5mm] table {Figures/Data/learned_fr_single.txt};
\addlegendentry{Learned filter, single ch. }
    
\end{axis}
\node at (0.7, 3.6) {(b)};
\end{tikzpicture}
    \vspace{-0.3cm}
    \caption{(a): SER performance versus $V_{\text{p}}$ for the single channel scenario, the blue curve corresponds to the baseline setup but without applying DPD; (b): Frequency response of the learned filters when the receiver MF roll-off factor is set to  $10\%$. The frequency response of the RRC filter with $10\%$ roll-off is also shown as a reference.
    }
    \label{fig:fig1}
    \vspace{-0.25cm}
\end{figure}
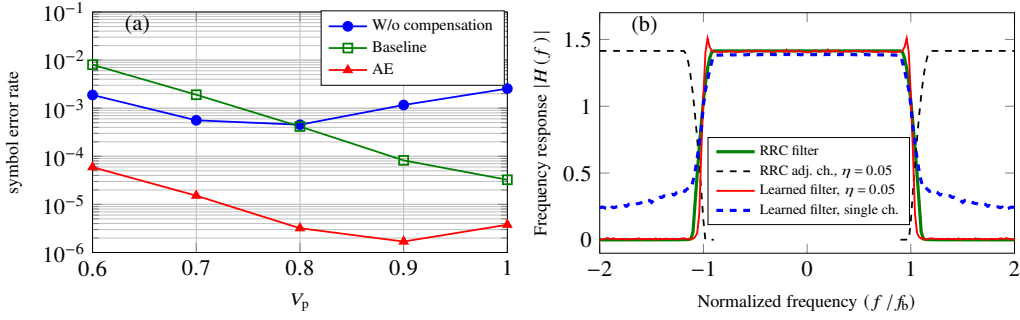

\noindent \emph{Impact of guard band $\eta$:}
We start by investigating a single-channel scenario.
Fig.~\ref{fig:fig1} (a) presents the SER of the proposed system when the receiver MF uses $10\%$ roll-off. For a range of considered $V_{\text{p}}$,
the proposed approach achieves significantly better performance than the considered baseline. However, by looking at the frequency response of the learned PS filter, as shown with the blue dashed curve in Fig.~\ref{fig:fig1} (b), we observe that compared to the RRC filter with $10\%$ roll-off, the learned filter has a significant amount of out-of-band (OOB) energy, 
which will introduce significant ICI with narrowly-spaced neighbouring channels and make it unsuitable for superchannel systems.
In contrast, when training the proposed AE in a 3-channel system with $\eta=0.05$, 
Fig.~\ref{fig:fig1} (b) shows that 
the learned filter restricts the OOB energy and 
has a narrower frequency response than the RRC filter, indicating that the trainable filter learns to limit ICI.

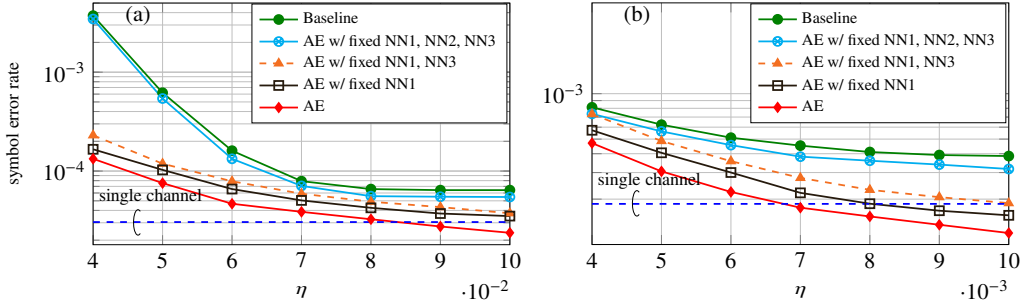
\begin{figure}[t]
    \centering
    \begin{tikzpicture}
\begin{semilogyaxis}[%
    scale only axis,
    tick label style={scaled ticks=base 10:2},
	xlabel= $\eta$ , ylabel= symbol error rate,
	xmin=0.04, xmax=0.1,%
	ymin=1.8e-5, ymax=5e-3,%
	grid=both,%
 	width=6.5cm, height=3.8cm,%
	legend cell align={left},%
	label style={font=\footnotesize},
	tick style={font=\footnotesize},
	legend style={at={(0.38,0.75)},anchor=west,nodes={scale=0.9, transform shape}, font=\footnotesize}%
]%

\addplot +[thick, color=green!50!black, solid, mark=*, mark options={scale=1,solid }] coordinates { 
 (0.04, 3.73e-3)
 (0.05,6.191e-04) (0.06,1.607e-04)
(0.07,7.904e-05) (0.08, 6.5724e-05) 
(0.09,6.4142e-5)
(0.1, 6.415e-05) 
};\addlegendentry{Baseline}

\addplot +[thick, color=cyan, solid, mark=otimes, mark options={scale=1,solid }] coordinates { 
 (0.04, 3.447e-3)
 (0.05,5.416e-04) (0.06,1.330e-04)
(0.07,7.078e-05) (0.08, 5.5724e-05) 
(0.09,5.5012e-5)
(0.1, 5.465e-05) 
};\addlegendentry{AE w/ fixed NN1, NN2, NN3}

\addplot +[thick, color=yellow!30!red, dashed, mark=triangle*, mark options={scale=1,solid }] coordinates { 
(0.04, 2.3e-4)
 (0.05,1.1878e-4) (0.06, 7.88e-5)
 (0.07,5.9e-05) (0.08, 4.8976e-05) 
(0.09, 4.32e-5) (0.1,3.745e-05) 
};\addlegendentry{AE w/ fixed NN1, NN3}

\addplot +[thick, color=brown!20!black, solid, mark=square, mark options={scale=1,solid }] coordinates { 
 (0.04, 1.66e-4)
 (0.05,1.0278e-4) (0.06, 6.576e-5)
 (0.07,5.05e-05) (0.08, 4.2476e-05) 
(0.09, 3.72e-5)(0.1,3.505e-05) 
};\addlegendentry{AE w/ fixed NN1}

\addplot +[thick, color=red, solid, mark=diamond*, mark options={scale=1,solid }] coordinates { 
(0.04,1.33e-4)
 (0.05,7.55e-5) (0.06, 4.67e-5)
 (0.07,3.87e-05) (0.08, 3.256e-05) 
(0.09, 2.75e-5)(0.1,2.375e-05) 
};\addlegendentry{AE}

\addplot +[thick, color=blue, dashed, mark=>, mark options={scale=1,solid }] coordinates { 
(0.04,3.05e-5) 
(0.1, 3.05e-5) 
};

\path[draw, thick, ->](0.055, 6.5724e-05) -- (0.08, 6.5724e-05);
\path[draw, thick, ->](6,-0.7) -- (6,-0.7);
	
\end{semilogyaxis}
\node at (0.7, 3.6) {(a)};
\node at (0.9, 0.75) { \footnotesize{single channel}};
\draw[black] (0.7, 0.35) [partial ellipse=-90+30+360:90-30:.07cm and 0.2cm];%
\end{tikzpicture}

\begin{tikzpicture}
\begin{semilogyaxis}[%
    scale only axis,
    tick label style={scaled ticks=base 10:3},
	xlabel= $\eta$, 
    ytick={1e-4,2e-4,3e-4,4e-4,5e-4, 6e-4, 7e-4, 8e-4, 9e-4, 1e-3},
    yticklabels={$ $,$ $,$ $,$ $,$ $, $ $, $ $,$ $,$ $, $10^{-3}$},
	xmin=0.004, xmax=0.01,%
	ymin=1.0e-4, ymax=4e-3,%
	grid=both,%
 	width=6.5cm, height=3.8cm,%
	legend cell align={left},%
	label style={font=\footnotesize},
	tick style={font=\footnotesize},
	legend style={at={(0.38,0.75)},anchor=west,nodes={scale=0.9, transform shape}, font=\footnotesize},
]%

\addplot +[thick, color=green!50!black, solid, mark=*, mark options={scale=1,solid }] coordinates { 
(0.004, 8.123e-4)
(0.005, 6.231e-4)
(0.006, 5.109e-4)
(0.007, 4.525e-4)
(0.008, 4.105e-04) 
(0.009, 3.920e-04)
(0.010, 3.860e-04) 
};\addlegendentry{Baseline}

\addplot +[thick, color=cyan, solid, mark=otimes, mark options={scale=1,solid }] coordinates { 
(0.004, 7.351e-4)
(0.005, 5.6231e-4)
(0.006, 4.561e-4)
(0.007, 3.825e-4)
(0.008, 3.595e-04) 
(0.009, 3.3880e-04) 
(0.010, 3.1680e-04) 
};\addlegendentry{AE w/ fixed NN1, NN2, NN3}

\addplot +[thick, color=yellow!30!red, dashed, mark=triangle*, mark options={scale=1,solid }] coordinates { 
 (0.004,7.323e-4)
 (0.005,4.856e-4)
 (0.006,3.569e-4)
 (0.007,2.765e-4)
 (0.008, 2.305e-4) 
 (0.009, 2.061e-4)
 (0.010, 1.881e-4)
};\addlegendentry{AE w/ fixed NN1, NN3}

\addplot +[thick, color=brown!20!black, solid, mark=square, mark options={scale=1,solid }] coordinates { 
 (0.004,5.723e-4)
 (0.005,4.056e-4)
 (0.006,3.0023e-4)
 (0.007,2.201e-4)
 (0.008, 1.865e-4) 
 (0.009, 1.676e-4)
 (0.010, 1.562e-4)
};\addlegendentry{AE w/ fixed NN1}

\addplot +[thick, color=red, solid, mark=diamond*, mark options={scale=1,solid }] coordinates { 
 (0.004,4.7023e-4)
 (0.005,3.055e-4)
 (0.006,2.232e-4)
 (0.007, 1.755e-4)
 (0.008, 1.535e-4) 
 (0.009, 1.352e-4)
 (0.010, 1.192e-4)
};\addlegendentry{AE}

\addplot +[thick,dashed, color=blue, mark=>, mark options={scale=1,solid }] coordinates { 
 (0.004,1.86e-4) 
(0.010,1.86e-4) 
}; 

\end{semilogyaxis}

\node at (0.7, 3.6) {(b)};
\node at (0.9, 1.0) { \footnotesize{single channel}};
\draw[black] (0.7, 0.65) [partial ellipse=-90+30+360:90-30:.07cm and 0.2cm];%
\end{tikzpicture}
    \vspace{-0.3cm}
    \caption{SER of the central channel of the proposed scheme and the baseline when the receiver MF roll-off factor set to (a) $10\%$ roll-off and (b) $1\%$ roll-off. The blue dashed curve corresponds to the SER of the baseline for a single channel scenario.}
    \label{fig:ser_wdm}
    \vspace{-0.7cm}
\end{figure}

\noindent \emph{SER analysis:}
We now evaluate the 
performance of the proposed system  setting the receiver MF roll-off factor to $10\%$ and $1\%$. The achieved SER 
for the central channel is shown in Fig.~\ref{fig:ser_wdm}.  As a reference, the SER performance of the baseline scheme is also shown. 
We remark that the clipping factor $V_\text{clip}$ and $V_{\text{p}}$ are optimized for the baseline scheme, while $V_{\text{p}}$ is set to $1$ in the proposed scheme for simplicity. Potentially, the performance of the proposed scheme can be further improved by optimizing $V_{\text{p}}$\textemdash  the optimal performance for the single channel case is achieved at $V_{\text{p}}=0.9$ (see Fig.~\ref{fig:fig1} (a)).
For roll-off factors of $10\%$ (Fig.~\ref{fig:ser_wdm}(a)) and $1\%$ (Fig.~\ref{fig:ser_wdm}(b)), the proposed  approach
outperforms
the baseline scheme over all considered guard bands. More importantly, compared to the baseline scheme, the guard band for the proposed scheme can be significantly reduced with limited  impact on the SER performance
\textemdash for the target SER where the baseline performance starts to saturate, the guard band can be reduced by around $37\%$ for $10\%$ roll-off  and around $50\%$ for $1\%$ roll-off.

\noindent \emph{Ablation study:}
In order to quantify the origin of the performance gains,
we carry out an ablation study by first freezing all the pre-trained NNs and individually unfreezing the NNs in the order of NN4, NN2, NN3, and NN1.
Fig.~\ref{fig:ser_wdm} shows that the SER performance of proposed system  improves every time one more NN is made trainable, indicating that the performance improvement of the proposed system can be attributed to the joint optimization of the mapper, the PS filter, the nonlinearity compensator (i.e., the DPD), and the de-mapper. 

\vspace{-0.1cm}
\section{Conclusion}
We have proposed an approach for high SE superchannel systems using an AE, where the transmitter network design follows the architecture of conventional systems. The resulting system achieves significantly better performance than 
the considered baseline scheme, and allows to increase the SE of superchannel systems by reducing the channel spacing without SER performance degradation. 

\vspace{0.2cm}
\noindent\footnotesize{\textbf{Acknowledgements: }This work was supported by the Knut and Alice Wallenberg Foundation, grant No.~2018.0090, and the Swedish Research Council under grant  No.~2018-0370.}

\vspace{-0.2cm}
\bibliographystyle{IEEEtran}
\bibliography{references}
\label{references}

\end{document}